\documentclass[twocolumn,showpacs,preprintnumbers,amsmath,amssymb]{revtex4}

\usepackage{graphicx}
\usepackage{dcolumn}
\usepackage{bm}

\begin{document}

\preprint{APS/123-QED}

\title{$\alpha$ cluster states in $^{44, 46, 52}$Ti} 

\author{M. Fukada,$^1$ M. K. Takimoto,$^2$ K. Ogino,$^3$  S. Ohkubo$^4$}

\affiliation{%
$^1$Kyoto Pharmaceutical University, Kyoto 607-8014, Japan\\
$^2$Department of Physics, Kyoto University, Kyoto 606-8502, Japan\\
$^3$Department of Nuclear Engineering, Kyoto University, Kyoto 606-8502, Japan\\
$^4$Department of Applied Science and Environment, Kochi Women's University, Kochi 780-8515, Japan
}%

\begin{abstract}
  $\alpha$ decaying states of $^{44, 46, 52}$Ti were investigated with angular correlation 
functions between $t$ and  $\alpha$ with the
 $^{40, 42, 48}$Ca($^7$Li, $t\alpha$)$^{40, 42, 48}$Ca reactions at E = 26.0 MeV.
Many $\alpha$ cluster states were newly observed in the 10 - 15 MeV excitation energy
 of $^{44}$Ti and their spin-parities were  assigned, in
 which $J^\pi=7^-$ state was found at 11.95 MeV as a candidate for the  member of 
the  $K=0_1^-$ negative parity band.
In $^{46}$Ti many $\alpha$ cluster states were also found in the  11 - 17 MeV excitation energy 
with the $^{42}$Ca($^7$Li, $t\alpha$)$^{42}$Ca reaction, though its strength is weak compared to $^{44}$Ti. 
No $\alpha$ cluster states were detected  for 
 the  $^{48}$Ca($^7$Li, $t\alpha$)$^{48}$Ca reaction, in which the number  of  
coincidence events decaying from $^{48}$Ca  was very small.  
\end{abstract}

\pacs{21.60.Gx, 25.70.-z, 27.40.+z}
\maketitle

\section{INTRODUCTION}
  
 The $\alpha$ cluster model  plays an important role in the study of  nuclear structure as well
 as the shell model and the collective model.
  It is especially important for understanding the  structure of light nuclei such as $^{8}$Be to $^{24}$Mg in the decay threshold energy region \cite{Horiuchi1972, Fujiwara1980}. 
To study $\alpha$ clusters  in heavier nuclei many theoretical and experimental investigations 
have  been performed focusing on $^{40}$Ca and $^{44}$Ti 
\cite{Ohkubo1998, Michel1998, Yamaya1998}. 
The negative parity $K=0_1^-$  band with   $\alpha$ cluster structure predicted by
  theoretical calculations \cite{Michel1986, Ohkubo1988, Ohkubo1998B}  
has been  discovered  using  $\alpha$ transfer experiments \cite{Yamaya1990, Yamaya1994}.

Detailed experimental studies have been devoted to  $^{44}$Ti, which is a typical 4N nucleus in
 the $fp$-shell, and the existence of $1^-$, $3^-$, $5^-$ states of  the $K=0_1^-$ band has been 
reconfirmed \cite{Yamaya1996, Guazzoni1993}.
It is interesting to explore the negative parity states with $J > 5$ of  the  $K=0_1^-$ band, but 
it seems  that the $^{40}$Ca($^{6}$Li, $d$)$^{44}$Ti reaction  is  not good at finding 
$\alpha$ cluster states at excitation energies above 10 MeV.
This is  because of the continuous spectrum
 of deuterons produced as a result of the disintegration of $^{6}$Li when bombarded  against  
$^{40}$Ca is large.
On the other hand  angle correlation reactions such as ($^{6}$Li,$d\alpha$) and 
($^{7}$Li,$t\alpha$) 
are effective for  investigating  the $\alpha$ cluster structure in that highly excited region, 
as we showed in the study of  the $\alpha$ cluster states in $^{38}$Ar in Ref.\cite{Fukada2005}.
Artemov \textit{et al}. \cite{Artemov1995} also studied the excited energy region above 11 MeV
 in  $^{44}$Ti  using the angle correlation reaction with the ($^{6}$Li,$d\alpha$) reaction 
but negative parity states with $J >5$ were not found.
 
On the other hand it has been shown that extra valence  neutrons play an important role 
in the stabilization of 
 the $\alpha$  cluster structure for non 4N nuclei.  
$^{46}$Ti is an analogue of $^{10}$Be and $^{22}$Ne  for which it has been shown
 that the  $\alpha$ cluster structure
 persists 
 \cite{Seya1981, Oertzen1997, Oertzen2006,Rogachev2001,Dufour2003,Kimura2007}.  
In $^{52}$Ti a possible $\alpha$ cluster structure was discussed for the ground band
 from the viewpoint of unified description of structure and  scattering  of the
 $\alpha$+$^{48}$Ca system \cite{Ohkubo1988B}. However, few experimental 
studies have been done in the highly excited energy region. 

In the present paper  the  $\alpha$ cluster structure in $^{44}$Ti,  $^{46}$Ti and  $^{52}$Ti were investigated with the ($^7$Li, $t\alpha$) reaction.

\section{EXPERIMENTAL PROCEDURE AND RESULTS}

 $^{7}$Li ions at 26 MeV incident energy from the Pelletron Accelerator at Kyoto University
 were bombarded  against  $^{40, 42, 48}$Ca targets.
The $^{40}$Ca (150 $\mu$g/cm$^2$) target was prepared by evaporating natural 
Ca metal ($^{40}$Ca component 96.9\%) on  carbon foil. The $^{42}$Ca 
(310 $\mu$g/cm$^2$) target was prepared by chemical vapor deposition from
 enriched (93.65\%) $^{42}$CaCO$_{3}$ on  carbon foil. 
The $^{48}$Ca (2.03 mg/cm$^2$) target was obtained by rolling enriched 
(97.80\%) $^{48}$Ca metal in  dried pure Ar gas.

Tritons were detected by the detector telescope  which consisted of a 150$\mu$m thick silicon
 $\Delta$E detector and a 2mm thick silicon E detector placed at 
 an  angle of $7.5^{\,\circ}$  to the  horizontal plane of the beam axis.
The acceptance angle of the telescope was 1.5$^\circ$ in the scattering plane 
and its solid angle was 2.0 msr.
A 120$\mu$m thick Al-foil was placed in front of the $\Delta$E detector to stop 
the $^7$Li ions which were elastically scattered by the target. 
The $\alpha$ particles were detected in coincidence with the tritons  by 
8 silicon photo-diode detectors placed at 8 angles
 between +$103.2^{\,\circ}$  and  +173.2$^{\,\circ}$. 
The aperture of each photo-diode was 10 mm width and 20 mm height.
Its solid angle was 15.1 msr and the depletion layer was 300 $\mu$m. 
A 100$\mu$m SSD was used to monitor the variation of the beam intensity and the target thickness.
 
Figure 1 and Figure 2 show the two dimensional energy spectra of $t-\alpha$ coincidences from the $^{40}$Ca target and $^{42}$Ca target,- respectively. In Fig.1 the locus (a) is from 
the $^{40}$Ca($^7$Li, $t\alpha$)$^{40}$Ca(g.s.) reaction and the locus (b) is from 
the $^{40}$Ca($^7$Li, $t\alpha$)$^{40}$Ca$^*$(3.35 MeV; $2_1^+$) reaction.
In Fig.2 the locus (a) is from the $^{42}$Ca($^7$Li, $t\alpha$)$^{42}$Ca(g.s.) reaction and 
the locus (b) is from the $^{42}$Ca($^7$Li, $t\alpha$)$^{42}$Ca$^*$(1.52 MeV; $2_1^+$) reaction.
In the case of  the $^{48}$Ca($^7$Li, $t\alpha$)$^{48}$Ca reaction, the locus of $t-\alpha$ 
that came out from $^{52}$Ti could not be discerned buried in the background data, though 
the thickness of the target was increased to 13.5 times and the beam integration of $^7$Li was 
increased  to a factor of 1.5 compared with $^{40}$Ca($^7$Li, $t\alpha$)$^{40}$Ca reaction.
Figure 3 shows the energy spectrum of the tritons obtained from the $^{40}$Ca($^7$Li, $t\alpha$)$^{40}$Ca(g.s.) reaction. Figure 4 shows the energy spectra of the tritons from $^{42}$Ca($^7$Li, $t\alpha$)$^{42}$Ca(g.s.) in the lower part and $^{42}$Ca($^7$Li,$t\alpha$)$^{42}$Ca($2_1^+$) reactions in the upper part. 
The horizontal axes show the excitation energy of $^{44}$Ti in Fig.~3 and $^{46}$Ti in Fig.~4, and the vertical axes show the summed counts of the tritons detected with eight silicon photo-diode detectors. 
The energy resolution of the triton was within 70 keV in both reactions.
The number  of $t-\alpha$ coincidence events decaying from $^{40}$Ca($2_1^+$) is very small compared to that from $^{40}$Ca(g.s.). In contrast
 the number of $t-\alpha$ coincidence events decaying from $^{42}$Ca($2_1^+$) is larger than that from $^{42}$Ca(g.s.) by a factor of 1.5.

\begin{figure}
\includegraphics[width=8.6cm,clip]{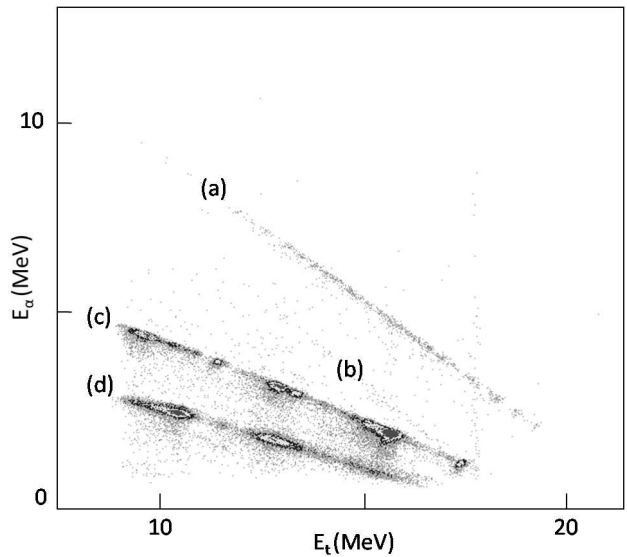}
\caption{\label{fig.1}  Two dimensional energy spectrum for the reaction  $^{40}$Ca($^7$Li, $t\alpha$)$^{40}$Ca. Dots represent $t-\alpha$ coincident events with the photo-diode at $+173.2^{\,\circ}$. (a) Locus from the $^{40}$Ca($^7$Li, $t\alpha$)$^{40}$Ca(g.s.) reaction. (b) Locus from the $^{40}$Ca($^7$Li, $t\alpha$)$^{40}$Ca$^*$(3.35 MeV; $2_1^+$) reaction. (c) Locus from the $^{16}$O($^7$Li, $t\alpha$)$^{16}$O(g.s.) reaction. (d) Locus from the $^{12}$C($^7$Li, $t\alpha$)$^{12}$C(g.s.) reaction.}
\end{figure}

\begin{figure}
\includegraphics[width=8.6cm,clip]{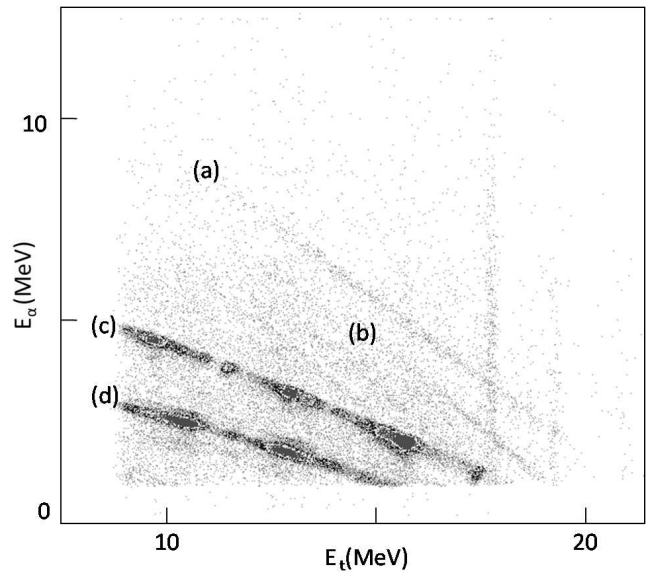}
\caption{\label{fig.2}  Two dimensional energy spectrum for the reaction  $^{42}$Ca($^7$Li, $t\alpha$)$^{42}$Ca. Dots represent $t-\alpha$ coincident events with the photo-diode at $+173.2^{\,\circ}$. (a) Locus from the $^{42}$Ca($^7$Li, $t\alpha$)$^{42}$Ca(g.s.) reaction. (b) Locus from the $^{42}$Ca($^7$Li, $t\alpha$)$^{42}$Ca$^*$(1.52; $2_1^+$) reaction. (c) Locus from the $^{16}$O($^7$Li, $t\alpha$)$^{16}$O(g.s.) reaction. (d) Locus from the $^{12}$C($^7$Li, $t\alpha$)$^{12}$C(g.s.) reaction.}
\end{figure}

\begin{figure*}
\includegraphics[bb=0 250 800 550, width=18cm,clip] {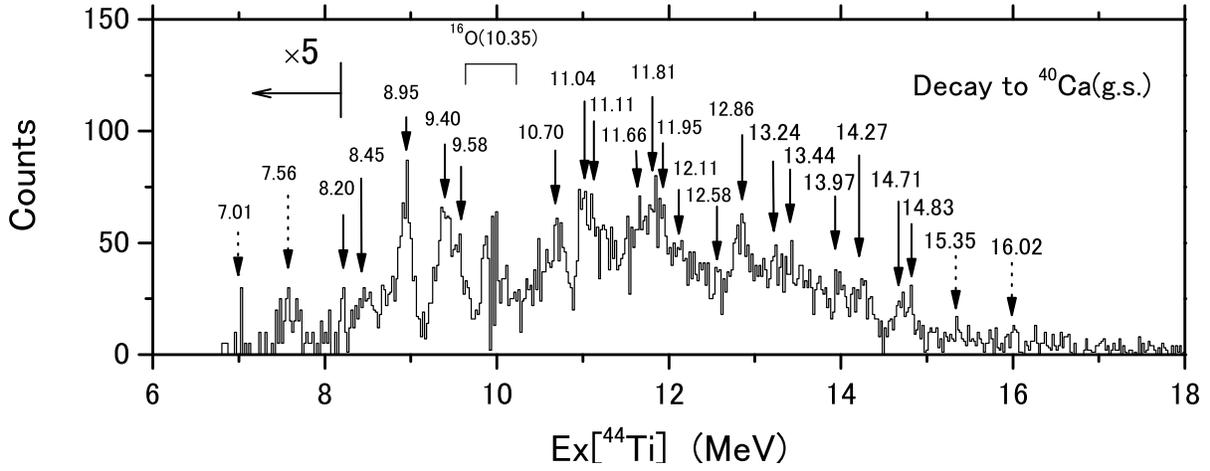}
\caption{\label{fig.3}  Spectrum of $t-\alpha$ coincidences from the $^{40}$Ca($^7$Li, $t\alpha$)$^{40}$Ca reaction. The counts are the sum of the events from the eight photo-diode detectors.
Peaks fitted with $|P_L(cos\theta)|^2$ are indicated by solid arrows. 
Data below 8.25 MeV in the excitation energy are multiplied by a factor of five.}
\end{figure*}

\begin{figure*}
\includegraphics[bb=0 50 800 550, width=18cm,clip] {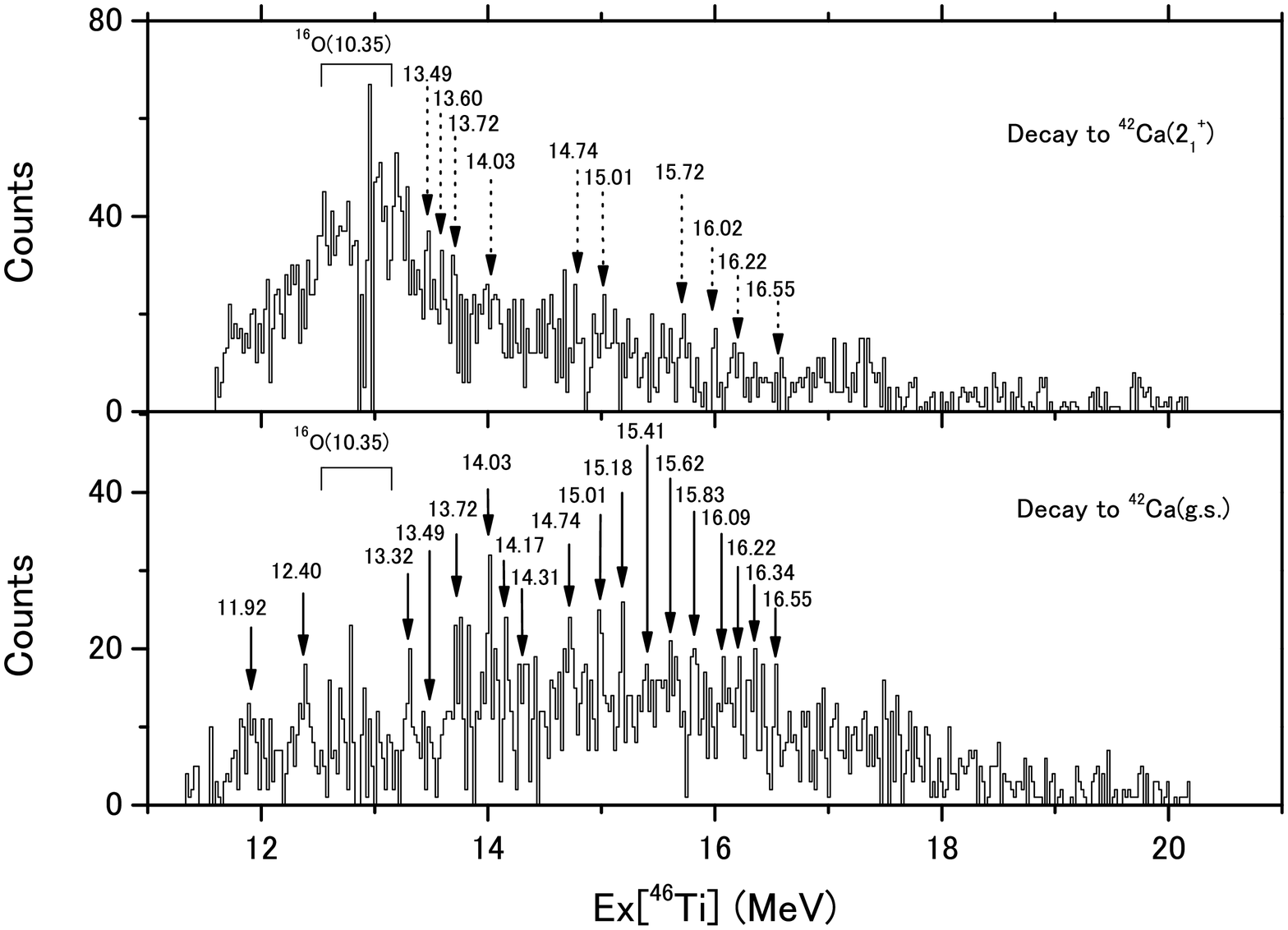}
\caption{\label{fig.4}  Spectra of $t-\alpha$ coincidences from the
 $^{42}$Ca($^7$Li, $t\alpha$)$^{42}$Ca reaction. The upper part is from that decaying to the
 first excited state (1.524 MeV, $2_1^+$) of $^{42}$Ca. The lower part is from that decaying to the ground state of $^{42}$Ca.  Both counts are the sum of the events from the eight photo-diode detectors.
Peaks fitted with $|P_L(cos\theta)|^2$ are indicated by solid arrows. }
\end{figure*}

\begin{figure}
\includegraphics[bb=-30 0 170 490 ,width=8.6cm,clip] {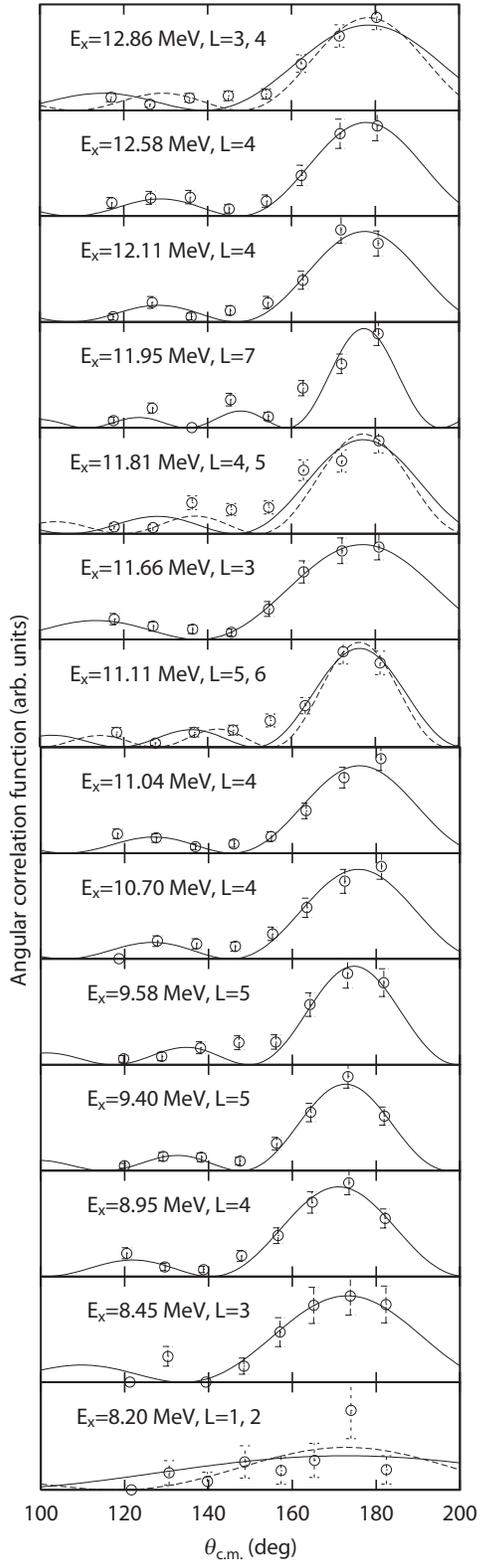}
\caption{\label{fig.5} Angular correlation functions at $E_{\rm x}$($^{44}$Ti)=8.20, 8.45, 8.95, 9.40, 9.58, 10.70, 11.04, 11.11, 11.66, 11.81, 11.95, 12.11, 12.58 and 12.86  MeV.
 The solid lines are the best $|P_L(cos\theta)|^2$ fits to the data.
The dashed lines show the second best fits in the case it is difficult to obtain unique L-values. }
\end{figure}

\begin{figure}
\includegraphics[bb=50 30 250 270, width=8.6cm,clip] {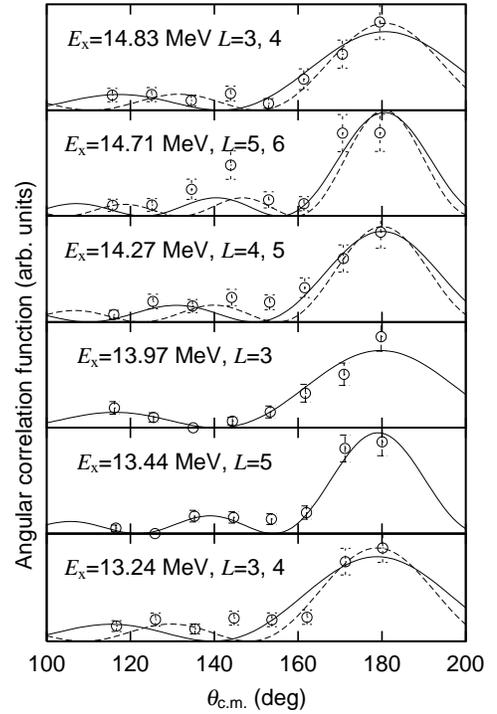}
\caption{\label{fig.6} Angular correlation functions at $E_{\rm x}$($^{44}$Ti)=13.24, 13.44, 13.97, 14.27, 14.71 and 14.83  MeV.
  The solid lines are the best $|P_L(cos\theta)|^2$ fits to the data.
The dashed lines show the second best fits in the case it is difficult to obtain unique L-values.}
\end{figure}

\begin{figure}
\includegraphics[bb=-30 0 170 490, width=8.6cm,clip] {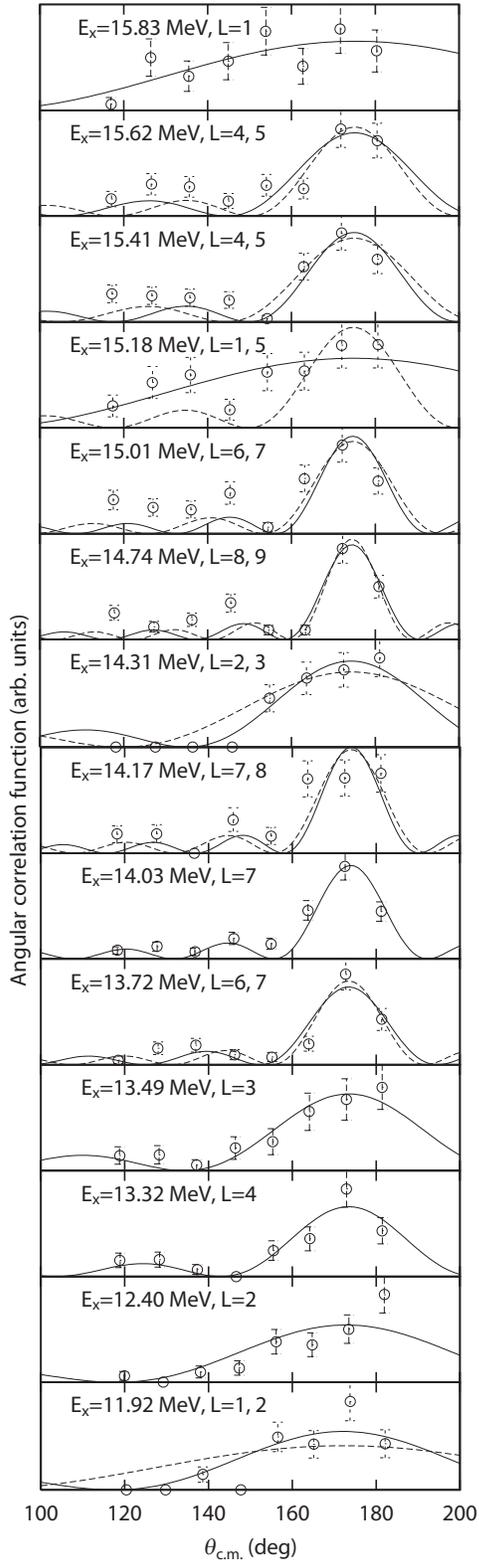}
\caption{\label{fig.7} Angular correlation functions at $E_{\rm x}$($^{46}$Ti)=11.92, 12.40, 13.32, 13.49, 13.72, 14.03, 14.17, 14.31, 14.74, 15.01, 15.18, 15.41, 15.62 and 15.83  MeV.
Data are from the $^{42}$Ca($^{7}$Li,$t\alpha$)$^{42}$Ti(g.s.) reaction.
 The solid lines are the best $|P_L(cos\theta)|^2$ fits to the data.
The dashed lines show the second best fits in the case it is difficult to obtain unique L-values.}
\end{figure}

\begin{figure}
\includegraphics[bb=50 30 250 200, width=8.6cm,clip] {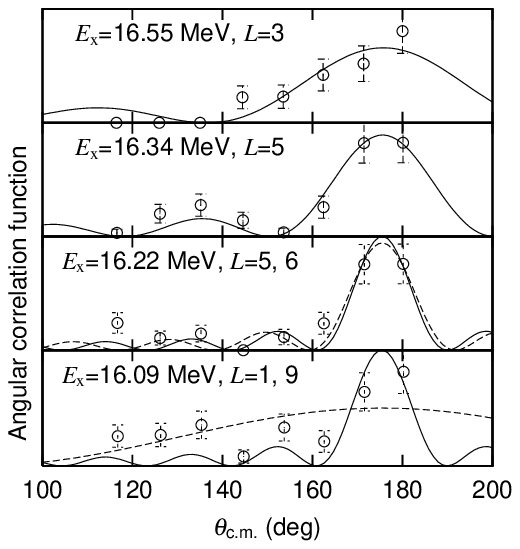}
\caption{\label{fig.8} Angular correlation functions at $E_{\rm x}$($^{46}$Ti)=16.09, 16.22, 16.34 and 16.55 MeV.
Data are from the $^{42}$Ca($^{7}$Li,$t\alpha$)$^{42}$Ti(g.s.) reaction.
 The solid lines are the best $|P_L(cos\theta)|^2$ fits to the data.
The dashed lines show the second best fits in the case it is difficult to obtain unique L-values.}
\end{figure}

\section{ANALYSIS AND DISCUSSION}

Figure 5 and Figure 6 show the angular correlation distributions of the levels excited 
in the $^{40}$Ca($^7$Li, $t\alpha$)$^{40}$Ca(g.s.) reaction.
 The experimental data have been  fitted with the squares of the Legendre polynomial $|P_L(cos\theta)|^2$. 
Figure 7 and Figure 8 show the angular correlation distributions of the levels excited in the $^{42}$Ca($^7$Li, $t\alpha$)$^{42}$Ca(g.s.) reaction.

In our present correlation experiment the detection angle of the triton was fixed 
at $\Theta_{lab}$ = 
7.5$^\circ$. When the triton is detected at angles other than 0$^\circ$, it is known that the 
angular distribution patterns of the $\alpha$ particle shift from 180$^\circ$ symmetry in the
 center of mass system in the reaction plane, because the wave function of the triton is distorted
 in the exit channel of the reaction. 
The amount of this angle shift tended to decrease with the increase of the excitation energy 
of the $\alpha$ emitting nucleus \cite{Cunsolo1980, Artemov1991}.
 In the present experiment,  the shift also decreased linearly from +7$^\circ$  as the excitation
 energy of  the $^{44,46}$Ti nuclei got higher.
In fitting our experimental data the amount of the shift calculated from that linear relationship
 was used. 

\subsection{$^{44}$Ti}

\begin{figure*}
\includegraphics[width=18cm,clip] {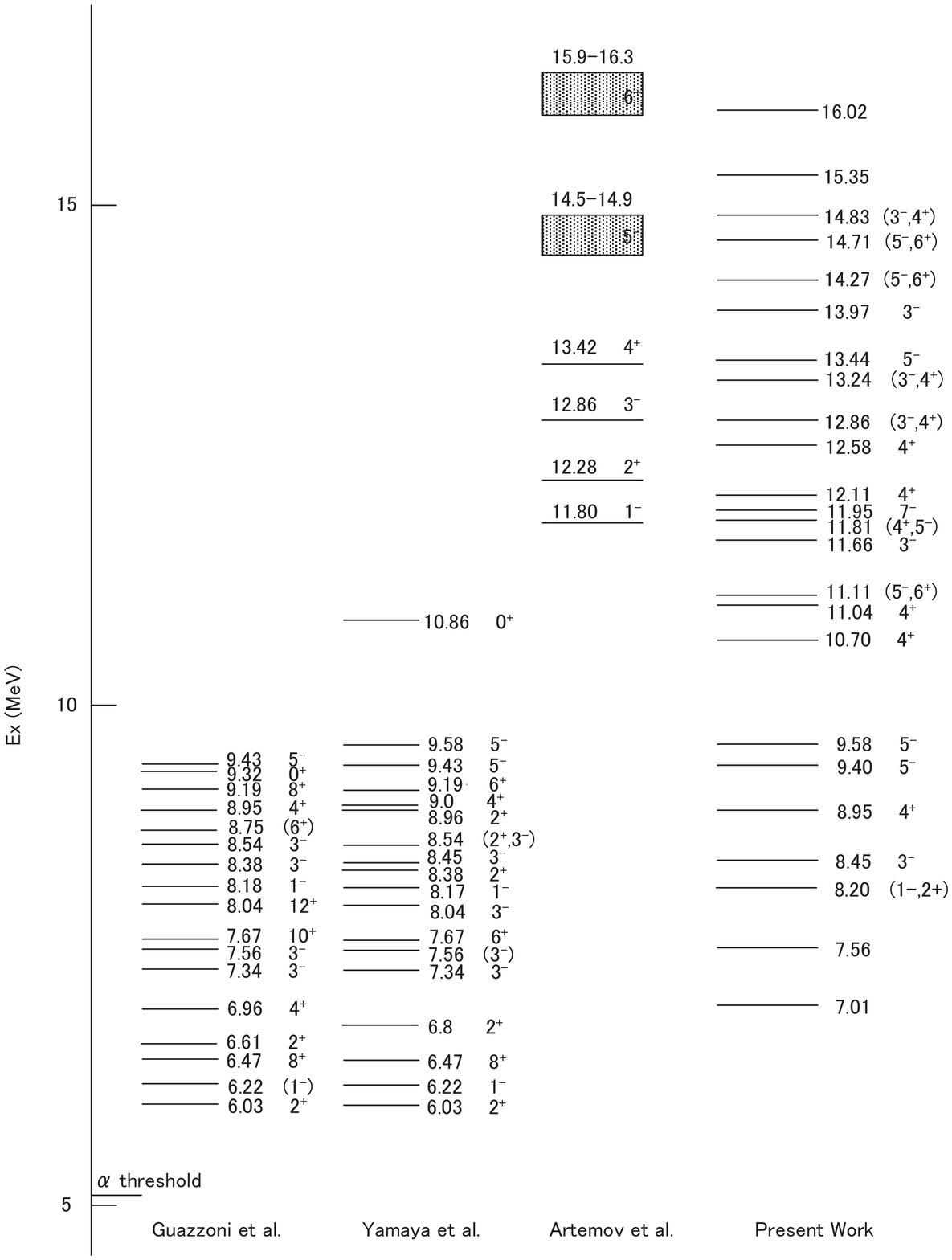}
\caption{\label{fig.9}  Energy levels of $^{44}$Ti observed in the $^{40}$Ca($^6$Li, $d$)$^{44}$Ti reaction by
 Guazzoni \textit{et al}. \cite{Guazzoni1993} as well as
 Yamaya \textit{et al}. \cite{Yamaya1990, Yamaya1996}, the $^{40}$Ca($^6$Li, $d\alpha$)$^{40}$Ca reaction by 
Artemov \textit{et al}. \cite{Artemov1995} and the present $^{40}$Ca($^7$Li, $t\alpha$)$^{40}$Ca reaction are shown. }
\end{figure*}

Figure 9 shows the energy levels of $^{44}$Ti above  $\alpha$ decay threshold that have been 
observed up to now  in $\alpha$ transfer reactions.
Yamaya \textit{et al}.  \cite{Yamaya1990, Yamaya1996}  used the ($^{6}$Li, $d$) reaction
 with incident energies of 50 MeV \cite{Yamaya1990} and 37 MeV \cite{Yamaya1996}.
Similarly, Guazzoni \textit{et al}. \cite{Guazzoni1993} used the ($^{6}$Li, $d$) reaction with incident 
energy of 60.1 MeV. 
In those experiments  no levels could be obtained above 11 MeV.
Meanwhile Artemov \textit{et al}.  \cite{Artemov1995}   performed the ($^{6}$Li,$d$$\alpha$)  
correlation experiment with an incident beam of 22 MeV and reported some $\alpha$ cluster 
states and bands in the excitation energy of 11 MeV to 16 MeV, though they did not mention
 about the structures below 11 MeV. 
We have found   more than twenty  $\alpha$ cluster states with the ($^{7}$Li, $t$$\alpha$) reaction
 and our results are compared to others' as follows.

7.01 MeV, 7.56 MeV: Because the threshold level of discriminators
 of the particle detecting system reduced the $\alpha$ yields, we could not obtain
 angular correlation distributions of these states. However these states show clear peaks as
 seen in Fig. 3, 
although there may be some levels scattered around 7.56 MeV. We concluded these states 
are $\alpha$ cluster states which correspond to the ones found by 
Yamaya \textit{et al}.  \cite{Yamaya1990, Yamaya1996}   and 
Guazzoni \textit{et al}. \cite{Guazzoni1993}.

8.20 MeV: This is the lowest excited state that  could be fitted with the angular correlation 
function and we assigned its spin-parity as  $J^\pi=1^-$ or $2^+$.
Yamaya \textit{et al}.  \cite{Yamaya1990, Yamaya1996}   and Guazzoni \textit{et al}. \cite{Guazzoni1993} 
 found $J^\pi=1^-$ state at 8.17 MeV and 8.18 MeV respectively.
 We may conclude the present state corresponds to those levels.

8.45 MeV: 
We could assign its spin-parity to $J^\pi=3^-$, though we detected a weak peak on the shoulder of this state. 
Meanwhile Yamaya \textit{et al}.  \cite{Yamaya1990, Yamaya1996}   assigned $J^\pi=3^-$ to this state and they also found 
the state with $J^\pi=2^+$ or $3^-$ at 8.54 MeV.
In addition Guazzoni \textit{et al}. \cite{Guazzoni1993} found $J^\pi=3^-$ state
 at 8.38 MeV and 8.54 MeV. 

8.95 MeV: This is a strongly activated state as seen in Fig.~3  and we assigned the spin-parity 
to $J^\pi=4^+$.
  Yamaya \textit{et al}. \cite{Yamaya1990, Yamaya1996}  found the 8.96 MeV state whose 
spin-parity was assigned to $J^\pi=2^+$, whereas 
Guazzoni \textit{et al}.  \cite{Guazzoni1993} found the 8.95 MeV state with $J^\pi=4^+$.

9.40 MeV: It is also a strongly excited state whose peak width is a little broadened and
 we assigned its spin-parity  as $J^\pi=5^-$. We could fit it with the $L=5$ angular correlation 
function very well, though it may consist of two levels, indicating that it is the same $5^-$ level 
at 9.43 MeV found by 
Yamaya \textit{et al}. \cite{Yamaya1990, Yamaya1996}  and
 Guazzoni \textit{et al}. \cite{Guazzoni1993}.

9.58 MeV: This state was well fitted with $L=5$ and assigned to be a  $J^\pi=5^-$ level. 
This level is in good agreement with Yamaya \textit{et al}. \cite{Yamaya1990, Yamaya1996}.

10.70 MeV: We assigned its spin-parity to be  $J^\pi=4^+$. 

11.04 MeV: It is a newly found level and we assigned its spin-parity to $J^\pi=4^+$. 

11.11 MeV, 11.66 MeV: We assigned the spin-parities of these states to
 $J^\pi=5^-$ or $6^+$,   and $J^\pi=3^-$,  respectively.

11.81 MeV: We assigned the spin of this state to $J^\pi=4^+$ or $5^-$ whereas 
Artemov {\it et al.} \cite{Artemov1995} assigned it to $J^\pi=1^-$.

11.95 MeV: This newly found state was assigned to be a $J^\pi=7^-$ level,
 because the phase pattern is more reproduced by  $L=7$, though $L=4$ improves the fit in the
 angles larger than 150$^{\circ}$.
It seems to correspond to the $7^-$ state of the $K=0_1^-$ band predicted around 12.4 MeV 
 in Ref. \cite{Michel1986}. 

12.11 MeV, 12.58 MeV: These state were well fitted with $L=4$ and assigned to $J^\pi=4^+$. 

12.86 MeV: This is guessed to be either $J^\pi=3^-$ or $4^+$. 
Artemov \textit{et al.} \cite{Artemov1995}  found a  $J^\pi=3^-$ state at 12.86 MeV, 
and the present state may correspond to that state. 

13.24 MeV: This is also guessed to be either $J^\pi=3^-$ or $4^+$. 

13.44 MeV: This state was well fitted with $L=5$ and we assigned its spin-parity to $J^\pi=5^-$. 
Whereas Artemov \textit{et al}.  \cite{Artemov1995}   reported the state of $J^\pi=4^+$ around
 13.42 MeV.

13.97 MeV: We assigned the spin-parity of this state to $J^\pi=3^-$. 
This state  corresponds well to the $J^\pi=3^-$ state of  the higher
 nodal $K=0_2^-$ band with the well-developed 
$\alpha$+$^{40}$Ca cluster structure   discussed  by 
Ohkubo \textit{et al}. \cite{Ohkubo1998B}

14.27 MeV: We assigned the spin-parity of this state to $J^\pi=4^+$ or $J^\pi=5^-$.
Because the state has a clear peak and it is apart from the $J^\pi=5^-$ state found 
by Artemov \textit{et al}. \cite{Artemov1995}, it can not correspond to the state with 
a wide energy width of 14.5$\sim$14.9 MeV.

14.71 MeV, 14.83 MeV: These states are guessed to be  $J^\pi=5^-$ or $6^+$,   and $J^\pi=3^-$ or $4^+$,  respectively.
The present levels lie in a broad band centering on the excitation energy of 14.7$\pm$0.2 MeV 
  for which Artemov {\it et al.} \cite{Artemov1995}
   presumed  $J^\pi=5^-$. 

15.35 MeV, 16.02 MeV: The yields at these states were too small to obtain $L$-values from
 the angular correlation distributions. 
The $J^\pi=6^+$ band at 15.9$\sim$16.3 MeV denoted by 
Artemov \textit{et al}. \cite{Artemov1995}
   is the only excited state which has been reported 
above 15 MeV.
The present states with narrow peaks may show that the band $J^\pi=6^+$ state is fragmented in this area if the report by
 Artemov \textit{et al}.  \cite{Artemov1995}
  is correct. 

\subsection{$^{46}$Ti}

It has been shown that $\alpha$  clustering persists in neutron rich nuclei, for example,
in  $^{10}$Be \cite{Oertzen1997,Oertzen2006} with the two  valence neutrons added to
 the $\alpha$+$\alpha$ cluster structure,
 in the  $0p$-shell region and in  $^{22}$Ne in which two  valence neutrons added to the 
$\alpha$+$^{16}$O cluster structure in the $sd$-shell region 
\cite{Oertzen2006,Rogachev2001,Dufour2003,Kimura2007}.
It is very interesting to study to what extent the $\alpha$  clustering persists 
in the $fp$-shell region when the extra valence  neutrons are added to the typical nucleus $^{44}$Ti
with the $\alpha$+$^{40}$Ca cluster structure.  $^{46}$Ti, for  which  few experimental
 and theoretical $\alpha$ cluster studies  have been devoted,  is an analog of  $^{22}$Ne.
Although many excited states of $^{46}$Ti have been reported below 10 MeV, the purposes of 
those experiments were not to investigate whether $^{46}$Ti shows the $^{42}$Ca+$\alpha$ 
structure.
An $\alpha$ transfer experiment with the ($^{6}$Li,$d$) reaction  was reported
 in Ref.\cite{Fulbright1977}. However,  only the ground 
state and the first excited state of $^{46}$Ti were examined.

\begin{table}[tbh]
  \caption[test]{ Excited states in $^{46}$Ti  with adopted  $J^\pi$ }
  \label{TABLE 1}
  \begin{center}
  \setlength{\tabcolsep}{3pt}
  \footnotesize
  \begin{tabular} {cccrcl} \hline \hline
	Decay mode & Excited Energy (MeV) &  $J^\pi$  \\ \hline
	g.s. & 11.92 &  $(1^-, 2^+)$ \\
	   & 12.40 & $2^+$  \\
	   & 13.32 & $4^+$ \\
	   & 13.49 & $3^-$ \\
	   & 13.72 & $(6^+, 7^-)$ \\
	   & 14.03 & $7^-$ \\
	   & 14.17 & $(7^-, 8^+)$ \\
	   & 14.31 & $(2^+, 3^-)$ \\
	   & 14.74 & $(8^+, 9^-)$ \\
	   & 15.01 & $(6^+, 7^-)$ \\
	   & 15.18 & $(1^-, 5^-)$ \\
	   & 15.41 & $(4^+, 5^-)$ \\
	   & 15.62 & $(4^+, 5^-)$ \\
	   & 15.83 & $1^-$ \\
	   & 16.09 & $(1^-, 9^-)$ \\
	   & 16.22 & $(5^-, 6^+)$ \\
	   & 16.34 & $5^-$ \\
	   & 16.55 & $3^-$ \\ \hline
	  
	$2_1^+$ & 13.49 &  \\
	 	& 13.60 &  \\
	 	& 13.72 &  \\
	 	& 14.03 &  \\
	 	& 14.74 &  \\
	 	& 15.01 &  \\
	 	& 15.72 &  \\
	 	& 16.02 &  \\
	 	& 16.22 &  \\
	 	& 16.55 &  \\
	 \hline \hline
  \end{tabular}
  \end{center}
\end{table}%

 The present  angular correlation experiment  is the first to investigate the $\alpha$ 
cluster structure of $^{46}$Ti. 
Table I shows the observed $\alpha$ cluster states in $^{46}$Ti  with the 
spin assignments obtained from the analysis given in Figs. 7 and Fig. 8.

 We may conclude that the present excited states with natural
 parities have  the $\alpha$+$^{42}$Ca(g.s.) or  $\alpha$+$^{42}$Ca(2$_1^+$) structure in
  $^{46}$Ti, although   $\alpha$ strengths are not as  strong as in $^{44}$Ti.  
On the other hand  the main  feature of the present $^{42}$Ca($^{7}$Li,$t\alpha$)$^{42}$Ca reaction 
is that the decay to the $2_1^+$ state of $^{42}$Ca was stronger than to the ground state
 unlike the $^{40}$Ca($^{7}$Li,$t\alpha$)$^{40}$Ca reaction.
Because  almost  the same number of states are excited in $^{44}$Ti and $^{46}$Ti 
in the relevant  energy region,
 $\alpha$ cluster structure in $^{46}$Ti  may be  analogous to   $^{44}$Ti. 
Since the core  $^{42}$Ca  is soft compared with the core $^{40}$Ca,
the two $\alpha$ cluster structures with the $\alpha$+$^{42}$Ca(g.s.)  and 
  $\alpha$+$^{42}$Ca(2$_1^+$) configurations may coexist in $^{46}$Ti  in the relevant energy region. 

\subsection{$^{52}$Ti}

 The $^{40}$Ca $\sim$ $^{48}$Ca nuclei are isotopes with  $Z$=20,  in which neutrons
 fill  the $0f_{7/2}$ shell as the mass number increases from $A$=40 to 48, until $^{48}$Ca
 becomes a doubly closed  nucleus. 
The persistency  of  $\alpha$ clustering in  such
  nuclei is  an interesting theme.
Although there are some experiments which have measured the ground band for
 $^{52}$Ti \cite{Fulbright1977, Mathiak1976, Morgan1977}, 
it is important to explore the cluster state 
in the higher excited energy region  using  the correlation method.

\begin{table}[tbh]
  \caption[test]{Relative ratio of the reaction cross sections}
  \label{TABLE 2}
  \begin{center}
  \setlength{\tabcolsep}{2pt}
  \footnotesize
  \begin{tabular} {ccccc} \hline \hline
	Reaction &  $\frac{d\sigma}{d\Omega } $ (arb. unit)  &   Relative ratio  \\ \hline
	$^{40}$Ca($^{7}$Li,$t\alpha$)$^{40}$Ca(g.s.) & 3.6 $\pm$0.04$\times$10$^{-3}$ & 1.0   \\
	$^{42}$Ca($^{7}$Li,$t\alpha$)$^{42}$Ca(g.s.)   & 2.02 $\pm$0.69$\times$10$^{-4}$ & 0.056 \\
	$^{42}$Ca($^{7}$Li,$t\alpha$)$^{42}$Ca($2_1^+$)  & 3.02 $\pm$0.84$\times$10$^{-4}$ & 0.084 \\
	$^{48}$Ca($^{7}$Li,$t\alpha$)$^{48}$Ca(g.s.) & $\ll 5.2 \times 10 ^ -6$  & $\ll 0.0014 $ \\
	\hline \hline
  \end{tabular}
  \end{center}
\end{table}%

To examine the
 $\alpha$ cluster structure of $^{52}$Ti,  we performed the 
$^{48}$Ca($^{7}$Li,t)$^{52}$Ti($\alpha$)$^{48}$Ca reaction experiment 
using thin $^{48}$Ca metal targets deposited onto thin carbon foils.
 However, any yield was not obtained in spite of a long machine time.
In the following experiment a thick $^{48}$Ca target made by 
pressing $^{48}$Ca metal was used.  Nevertheless, the cross section of the $\alpha$ cluster transfer
 reaction that produced  $^{52}$Ti was extremely small and no excited states were obtained. 
In Table II the variation of the relative cross sections in the ($^{7}$Li,t$\alpha)$ reaction 
targeting on  $^{40}$Ca, $^{42}$Ca and $^{48}$Ca is shown. 
The cross sections there correspond to the coincidence events between the triton detectors in
 $\Theta_L=7.5^\circ$ and the eight $\alpha$ detectors, and not the total cross section. 
The strengths of 5.6 \% in the $^{42}$Ca($^{7}$Li,$t\alpha$)$^{42}$Ca(g.s.) reaction and 
8.4 \% in the $^{42}$Ca($^{7}$Li,$t\alpha$)$^{42}$Ca($2_1^+$) reaction were obtained 
 when the strength in  the $^{40} $Ca($^{7}$Li,$t\alpha$)$^{40}$Ca(g.s.) reaction was assumed 
to be 1.0. 
In contrast the strength in $^{48}$Ca($^{7}$Li,$t\alpha$)$^{48}$Ca reaction was only 
0.14 \% or less.
This suggests that the $\alpha$  clustering is the  more suppressed  in $^{52}$Ti   nuclei  the more 
extra neutrons are added filling the $0f_{7/2}$ shell.

\section{CONCLUSION}

Angular correlation experiments with the ($^{7}$Li,$t\alpha$) reaction were performed
 using  $^{40}$Ca, $^{42}$Ca and $^{48}$Ca targets to investigate the $\alpha$ cluster states 
of   $^{44} $Ti, $^{46}$Ti  and $^{52}$Ti nuclei.

 For $^{44}$Ti  twenty-four $\alpha$ cluster states were observed in the excitation energy 
of 7 MeV to 16 MeV. 
We could uniquely assign spin-parities to the eight states at 10.70 MeV ($4^+$),
11.04 MeV ($4^+$), 11.66 MeV ($3^-$), 11.95 MeV ($7^-$), 12.11 MeV ($4^+$), 12.58 MeV ($4^+$)
13.44 MeV ($5^-$) and 13.97 MeV ($3^-$) as newly found $\alpha$ cluster levels, in which the state at 
11.95 MeV ($7^-$) seems  correspond to the $J^\pi=7^-$ state of the $K=0_1^-$ band predicted
 in Ref.\cite{Michel1986} but not observed
 in the ($^6$Li,d) $\alpha$  transfer reactions
 \cite{Yamaya1990,Yamaya1996,Guazzoni1993}.  
The seven states at 11.11 MeV ($5^-,6^+$), 11.81($4^+,5^-$), 
12.86 MeV ($3^-,4^+$), 13.24 MeV ($3^-,4^+$), 14.27 MeV ($4^+,5^-$), 14.71 MeV  ($5^-,6^+$),
 and 14.83 MeV ($3^-,4^+$) are also
 $\alpha$ cluster levels that had not been discovered in other studies, though some 
uncertainties remain for the spin assignments in the present study.
The state at 8.20 MeV  ($1^-,2^+$) may correspond to the  ($1^-$)  states at 8.17 MeV and 
8.18 MeV reported by other exexperiments \cite{Yamaya1990,Yamaya1996, Guazzoni1993}.
The states at 8.45 MeV ($J^\pi=3^-$), 8.95 MeV ($4^+$), 9.40 MeV  ($5^-$) and 
9.58 MeV ($5^-$)  are well agreement to the state at 8.45 MeV, 8.96 MeV,  9.43 MeV and
9.58 MeV reported by Yamaya \textit{et al.} \cite{Yamaya1990,Yamaya1996}.
 Spins could not be assigned to the two states at 7.01 MeV, 7.56 MeV
in lower excitation energy and the two states at 15.35 MeV, 16.02 MeV in higher excitation
 energy,  because the number  of  coincidence events was  too small to obtain $L$-values with
 the angular correlation functions.

For $^{46}$Ti  many candidates of the $\alpha$ cluster state are found in the excitation energy 
of 11 MeV to 17 MeV by the reactions that decay 
to the ground state and the first excited state of $^{42}$Ca.
Spin-parities of the seven states were uniquely assigned at 12.40 MeV ($2^+$), 13.32 MeV ($4^+$), 
13.49 MeV ($3^-$), 14.03 MeV ($7^-$) , 15.83 MeV ($1^-$), 16.34 MeV ($5^-$)  and 16.55 MeV ($3^-$).
We also assigned with some ambiguities to the eleven states at 11.92 MeV ($1^-,2^+$), 
13.72 MeV ($6^+,7^-$), 14.17 MeV ($7^-,8^+$),
14.31 MeV ($2^+,3^-$), 14.74 MeV ($8^+,9^-$), 15.01 MeV ($6^+,7^-$), 15.18 MeV ($1^-,5^-$),
15.41 MeV ($4^+,5^-$), 15.62 MeV ($4^+,5^-$), 16.09 MeV ($1^-,9^-$) and 16.22 MeV ($5^-,6^+$).

For $^{52}$Ti  no peaks could be detected in the energy spectrum of $t-\alpha$ coincidences
 from the $^{48}$Ca($^7$Li, $t\alpha$)$^{48}$Ca reaction due to very scare coincidence events. 

The variation of the ($^7$Li, $t\alpha$) reaction cross section with the increase of neutrons
 in  the  Ca  target was investigated with three Ca isotopes.
The relative $\alpha$ strength for $^{42}$Ca target was about one-tenth of $^{40}$Ca target
 and in the case of $^{48}$Ca its upper limit was 0.14 percent. 
This shows that the increase of neutrons weakens the structure of the $\alpha$ cluster
 in $^{52}$Ti nuclei.

\section{ACKNOWLEDGEMENT}

   The authors thank Professor K. Imai,  Professor H. Sakaguchi and other staff of the Pelletron Accelerator at Kyoto 
University for their encouragement and assistance in the operation of the accelerator.

\end{document}